\documentclass[12pt]{iopart}

\usepackage{iopams}  
\usepackage{caption}
\usepackage{xcolor}
\definecolor{light-gray}{gray}{0.96} 
\usepackage{listings}
\usepackage{graphicx}
\usepackage{placeins}
\usepackage{amsmath}
\usepackage{hyperref}

\definecolor{maroon}{cmyk}{0, 0.87, 0.68, 0.32}
\definecolor{halfgray}{gray}{0.55}
\definecolor{ipython_frame}{RGB}{207, 207, 207}
\definecolor{ipython_bg}{RGB}{247, 247, 247}
\definecolor{ipython_red}{RGB}{186, 33, 33}
\definecolor{ipython_green}{RGB}{0, 128, 0}
\definecolor{ipython_cyan}{RGB}{64, 128, 128}
\definecolor{ipython_purple}{RGB}{170, 34, 255}

\lstdefinelanguage{iPython}{
    morekeywords={access,and,break,class,continue,def,del,elif,else,except,exec,finally,for,from,global,if,import,in,is,lambda,not,or,pass,print,raise,return,try,while},%
    morekeywords=[2]{abs,all,any,basestring,bin,bool,bytearray,callable,chr,classmethod,cmp,compile,complex,delattr,dict,dir,divmod,enumerate,eval,execfile,file,filter,float,format,frozenset,getattr,globals,hasattr,hash,help,hex,id,input,int,isinstance,issubclass,iter,len,list,locals,long,map,max,memoryview,min,next,object,oct,open,ord,pow,property,range,raw_input,reduce,reload,repr,reversed,round,set,setattr,slice,sorted,staticmethod,str,sum,super,tuple,type,unichr,unicode,vars,xrange,zip,apply,buffer,coerce,intern},%
    sensitive=true,%
    morecomment=[l]\#,%
    morestring=[b]',%
    morestring=[b]",%
    morestring=[s]{'''}{'''},
    morestring=[s]{"""}{"""},
    morestring=[s]{r'}{'},
    morestring=[s]{r"}{"},%
    morestring=[s]{r'''}{'''},%
    morestring=[s]{r"""}{"""},%
    morestring=[s]{u'}{'},
    morestring=[s]{u"}{"},%
    morestring=[s]{u'''}{'''},%
    morestring=[s]{u"""}{"""},%
    literate=
    {á}{{\'a}}1 {é}{{\'e}}1 {í}{{\'i}}1 {ó}{{\'o}}1 {ú}{{\'u}}1
    {Á}{{\'A}}1 {É}{{\'E}}1 {Í}{{\'I}}1 {Ó}{{\'O}}1 {Ú}{{\'U}}1
    {à}{{\`a}}1 {è}{{\`e}}1 {ì}{{\`i}}1 {ò}{{\`o}}1 {ù}{{\`u}}1
    {À}{{\`A}}1 {È}{{\'E}}1 {Ì}{{\`I}}1 {Ò}{{\`O}}1 {Ù}{{\`U}}1
    {ä}{{\"a}}1 {ë}{{\"e}}1 {ï}{{\"i}}1 {ö}{{\"o}}1 {ü}{{\"u}}1
    {Ä}{{\"A}}1 {Ë}{{\"E}}1 {Ï}{{\"I}}1 {Ö}{{\"O}}1 {Ü}{{\"U}}1
    {â}{{\^a}}1 {ê}{{\^e}}1 {î}{{\^i}}1 {ô}{{\^o}}1 {û}{{\^u}}1
    {Â}{{\^A}}1 {Ê}{{\^E}}1 {Î}{{\^I}}1 {Ô}{{\^O}}1 {Û}{{\^U}}1
    {œ}{{\oe}}1 {Œ}{{\OE}}1 {æ}{{\ae}}1 {Æ}{{\AE}}1 {ß}{{\ss}}1
    {ç}{{\c c}}1 {Ç}{{\c C}}1 {ø}{{\o}}1 {å}{{\r a}}1 {Å}{{\r A}}1
    {€}{{\EUR}}1 {£}{{\pounds}}1
    {^}{{{\color{ipython_purple}\^{}}}}1
    {=}{{{\color{ipython_purple}=}}}1
    {+}{{{\color{ipython_purple}+}}}1
    {*}{{{\color{ipython_purple}$^\ast$}}}1
    {/}{{{\color{ipython_purple}/}}}1
    {+=}{{{+=}}}1
    {-=}{{{-=}}}1
    {*=}{{{$^\ast$=}}}1
    {/=}{{{/=}}}1,
    literate=
    *{-}{{{\color{ipython_purple}-}}}1
     {?}{{{\color{ipython_purple}?}}}1,
    identifierstyle=\color{black}\ttfamily,
    commentstyle=\color{ipython_cyan}\ttfamily,
    stringstyle=\color{ipython_red}\ttfamily,
    keepspaces=true,
    showspaces=false,
    showstringspaces=false,
    rulecolor=\color{ipython_frame},
    frame=single,
    frameround={t}{t}{t}{t},
    framexleftmargin=6mm,
    numbers=left,
    numberstyle=\tiny\color{halfgray},
    backgroundcolor=\color{ipython_bg},
    basicstyle=\scriptsize,
    keywordstyle=\color{ipython_green}\ttfamily,
}

\definecolor{codegreen}{rgb}{0,0.6,0}
\definecolor{codegray}{rgb}{0.5,0.5,0.5}
\definecolor{codepurple}{rgb}{0.58,0,0.82}
\definecolor{mygreen}{RGB}{28,172,0} 
\definecolor{mylilas}{RGB}{170,55,241}
\definecolor{backcolour}{rgb}{0.95,0.95,0.92}

\begin{document}

\title[TIGRE v3]{TIGRE v3: Efficient and easy to use iterative computed tomographic reconstruction toolbox for real datasets}

\author{Ander Biguri\textsuperscript{1}, Tomoyuki Sadakane\textsuperscript{2}, Reuben Lindroos, Yi Liu\textsuperscript{3}, Malena Sabat\'e Landman\textsuperscript{4}, Yi Du\textsuperscript{5,6}, Manasavee Lohvithee\textsuperscript{7}, Stefanie Kaser\textsuperscript{8},  Sepideh Hatamikia\textsuperscript{9,10}, Robert Bryll\textsuperscript{11}, Emilien Valat\textsuperscript{12}, Sarinrat Wonglee\textsuperscript{13}, Thomas Blumensath\textsuperscript{14}, Carola-Bibiane Sch{\"o}nlieb\textsuperscript{1}}

\address{\textsuperscript{1}Department of Applied Mathematics and Theoretical Physics (DAMTP), University of Cambridge, Cambridge, UK}
\address{\textsuperscript{2} J. Morita Manufacturing Corp., Japan}
\address{\textsuperscript{3}X-ray Imaging, Sensing and Sorting, CSIRO Mineral Resources, Sydney, Australia}
\address{\textsuperscript{4}Department of Mathematics, Emory University, Atlanta, USA}
\address{\textsuperscript{5}Key Laboratory of Carcinogenesis and Translational Research (Ministry of Education/Beijing), Department of Radiation Oncology, Peking University Cancer Hospital \& Institute, China}
\address{\textsuperscript{6}Institute of Medical Technology, Peking University Health Science Center, China.}
\address{\textsuperscript{7}Department of Nuclear Engineering, Chulalongkorn University,  Bangkok, Thailand } 
\address{\textsuperscript{8}Institute of High Energy Physics, Austrian Academy of Sciences, Vienna, Austria\footnote{Now at Institute for Biomedical Engineering, ETH Z\"urich  and University of Z\"urich, Switzerland}}
\address{\textsuperscript{9}Research Center for Clinical Al-Research in Omics and Medical Data Science
(CAROM), Department of Medicine, Danube Private University (DPU), Krems, Austria}
\address{\textsuperscript{10}Austrian Center for Medical Innovation and Technology (ACMIT),Wiener Neustadt, Austria}
\address{\textsuperscript{11}Mitutoyo Research \& Development America Inc., Kirkland, USA}
\address{\textsuperscript{12}Department of Mathematics, KTH Royal Institute of Technology, Stockholm, Sweden}
\address{\textsuperscript{13}Nuclear Research and Development Center, 
Thailand Institute of Nuclear Technology (TINT), Nakhon Nayok, Thailand}
\address{\textsuperscript{14}Institute of Sound and Vibration Research (ISVR), University of Southampton, Southampton, UK}

\ead{ander.biguri@gmail.com} 
\vspace{10pt}
\begin{indented}
\item[]February 2024
\end{indented}

\begin{abstract}
Computed Tomography (CT) has been widely adopted in medicine and it is increasingly being used in scientific and industrial applications. Parallelly, research in different mathematical areas concerning 
discrete inverse problems has led to the development of new sophisticated numerical solvers that can be applied in the context of CT. The Tomographic Iterative GPU-based Reconstruction (TIGRE) toolbox was born almost a decade ago precisely in the gap between mathematics and high performance computing for real CT data, providing user-friendly open-source 
software tools for image reconstruction. However, since its inception, the tools' features and codebase have had over a twenty-fold increase, and are now including greater geometric flexibility, a variety of modern algorithms for image reconstruction, high-performance computing features and support for other CT modalities, like proton CT. 
The purpose of this work is two-fold: first, it provides a structured overview of the current version of the TIGRE toolbox, providing appropriate descriptions 
and references 
, and 
serving as a comprehensive and peer-reviewed guide for the user; second, it is an opportunity to illustrate the performance of several of the available solvers showcasing real CT acquisitions, which are typically not be openly available to algorithm developers. 


\end{abstract}

%

\section{Introduction}

In the last years, tomography has emerged as an indispensable tool for imaging the inside of very different objects of interest. This non-destructive technique is considered safe as long as only small scanning doses are used, even though high doses can harm biological samples (including humans). For this reason, tomography is widely used by engineers, biologists, geophysicists, material and other scientist and medical doctors alike. 
The main challenge associated to tomographic imaging is that it is not a direct measuring technique, so we need to use tomographic reconstruction algorithms and these can produce images with very different properties and qualities, see, e.g. \cite{SIRT, doi:10.1137/1.9781611976670}. 

From all the existing types of tomography, Computed Tomography (CT) stands out as the most widely employed in medicine, science and engineering. The mathematical model corresponding to this kind of tomographic reconstruction has two main properties that make it challenging to solve, so decades after the inception of tomography 
it still remains of high interest for the discrete inverse problems mathematical community. First, the CT reconstruction problem is ill-posed: this means that, in the limited measurements regime and in the presence of noise, the reconstructions are very sensitive to small perturbations in the measurements, see, e.g. \cite{doi:10.1137/1.9780898718836, SIRT}. Moreover, these are scenarios of capital interest: in other words, the problem becomes more ill-posed when one reduces radiation doses in medical CT, reduces scanning time in industrial CT, or when physical limitations reduce the amount of available measuring angles. Second, modern applications tend to require bigger and bigger reconstructions either due to an increase in resolution, in object size, or in the amount of measurements \cite{doi:10.1137/1.9781611972344}. 

The constant need for efficient algorithms that can support both a trend towards more ill-posed and higher dimensional CT problems feeds on an ever-evolving paradigm of model reconstruction algorithms. To improve the image reconstruction quality, these typically leverage the noisy measurements with different kinds of prior information about the reconstruction. The most widely studied algorithms range from analytical methods, algebraic row-action methods, variational model-based methods or, more recently, machine learning based reconstruction methods. 
However, the very different nature of expertise required to develop and implement these algorithms has led to many scientist and medical doctors working with tomographic machines to only rely on direct classical solvers, 
despite ample evidence of the measurable improvements that iterative algorithms can achieve in the reconstruction pipeline \cite{ITERATIVES_GREAT_MEDICAL,ITERATIVES_GREAT_MEDICAL2}. On the other hand, researchers focusing on the development of algorithms for discrete inverse problems often lack the expertise and time to build software that is both efficient and robust enough for a lab-environment. This leads to a general norm of users optimizing scanning procedures for the closed proprietary software that the manufacturer of the scanners provide. More specifically, this software more often than not considers only a direct reconstruction algorithm, (e.g. filtered backprojection, or FBP) and requires high quality, high X-ray dose, over-sampled data. While this is available in some applications, shorter scanner time and doses are of high value in tomography, 
and therefore there is a concrete need for more user-friendly implementations of advanced mathematical tools to produce better image reconstructions in these scenarios. 

The Tomographic Iterative GPU-based Reconstruction (TIGRE) toolbox~\cite{TIGRE} was born exactly under this paradigm as an open source translational research tool to serve both algorithm developers (applied mathematicians and computer scientist without a software engineering background) and applied scientist working with real CT data. In this spirit, the TIGRE toolbox offers an easy to install platform where one can both reconstruct real data using advanced algorithms (with only a very basic level of mathematics and software knowledge) or write new mathematical methods for reconstruction and test them on high quality realistic test problems (with little knowledge of data managing pipelines or GPU-acceleration). Moreover, these new implemented and tested algorithms can be afterwards appended to the TIGRE toolbox to make them available to researchers wanting to try new algorithms on their data. Since its first release, the TIGRE toolbox has indeed become a community platform, used by researchers and scientists across the field of tomography. Indeed, since its publication the codebase of the toolbox has grown in over an order of magnitude, and, to the date, it has been cited in over 300 papers.

It is crucial to emphasize the significance of open source software (OSS) within academia, particularly in an era characterized by ubiquitous computing. In such a landscape, the advancement of science often needs highly specialized multidisciplinary skills spanning mathematics, computer science, engineering and physics, just to name a few. 
In order to create reproducible and reliable research, there is a need for free and transparent unified frameworks that offer a space for different research communities to meet and promote knowledge-exchange. 
The TIGRE toolbox contributes and enriches the increasingly wide ecosystem of OSS available for inverse problems in computed tomography, which differ in their approach to software development and distribution to applied scientists.
For general X-ray tomography, other GPU-based OSS exist, such as the widely used ASTRA toolbox~\cite{ASTRA} and other software that uses it. This is for example the case of the PyTorch~\cite{NEURIPS2019_bdbca288} compatible Tomosipo~\cite{tomosipo}, or the CIL toolbox~\cite{CIL} which also strongly focuses on the \textmu CT community. The latter uses both ASTRA and TIGRE as GPU projectors, yet providing several user-friendly tools to easily craft new iterative algorithms suited to the relevant experiments. On the other spectrum of target users, RTK~\cite{RTK} provides reconstruction code for data associated to medical scanners. Lastly, the TomoPy toolbox~\cite{tomopy} provides similar tools to the rest of the mentioned packages (including ASTRA wrappers), but was initially designed for synchrotron X-ray tomography. Recently, LEAP~\cite{LEAP} was also released, with auto-differentiable operators. Software focused on other types of tomography is also available. For example EIDORS~\cite{EIDORS} or pyEIT~\cite{pyeit} for Electrical Impedance Tomography, STIR~\cite{STIR} for emission tomography (such as Positron Emission Tomography and Single Particle Emission Computed Tomography), and SIRF~\cite{SIRF} or CASToR~\cite{merlin2018castor} for multi-modal tomography. Finally, other OSS that are mostly used by the discrete inverse problems (mathematics) community can be used for tomography ODL~\cite{odl}, AIRtools~\cite{hansen2018air} and IRtools~\cite{gazzola2019ir}, among others. However, these packages sometimes lack the required efficiency to support high dimensional realistic CT test problems. This is not an exhaustive list of the available software for tomography reconstruction. Each piece of software has nuanced differences in use cases, programming languages, licensing and support, which are out of the scope of this work to discuss. 

This work focuses on the TIGRE toolbox most current version 3.0, offering a structured overview of the toolbox including appropriate descriptions and references, and highlighting some of the most important features added since its first release. Subsequently, this paper aims to serve as a comprehensive and peer-reviewed guide for the user. Moreover, this work also aims 
to illustrate the performance of several of the solvers that are available on the TIGRE toolbox on real CT
acquisitions.

The article is structured as follows. Section \ref{Sec:methods} briefly introduces the mathematical background of iterative reconstruction algorithms, focusing on the algorithms currently implemented in the TIGRE toolbox. In this section, a structured description of the design of the TIGRE toolbox is also included, as well as highlights of the several new important features of the software such as novelties in the implementation, geometric flexibility and multi-GPU support. Next, Section \ref{Sec:results} presents five different examples based on real datasets, alongside the code needed to reproduce them. These correspond to a clinical scenario with patient data, a synchrotron parallel beam tomographic reconstruction, a Monte Carlo simulation of a proton CT problem, a neutron tomographic acquisition and an industrial \textmu CT dataset. Finally, some conclusions are given in Section \ref{Sec:conclusions}. A list of all the solvers provided by the TIGRE toolbox in its current state can be found in  \ref{Appendix1}.

\section{Methods}\label{Sec:methods}
This section provides both a context for the TIGRE toolbox and a description of the most important features that have been added since it was first published, almost a decade since this article. First, a brief description of the tomography reconstruction problem is given, along with a (non-exhaustive) review of iterative solvers. Afterwards, a brief description of the software architecture is included with the relevant references. Last, this section describes all the significant features that have been added to the toolbox, giving an in-depth description for those that have not been published elsewhere. These are of very different nature and have been added in the following order:
\begin{itemize}
    \item newly implemented mathematical optimization algorithms,
    \item new Python codebase implementation,
    \item new features concerning the applicability of the codes to real data: 
    \begin{itemize}
        \item increased geometric flexibility of the operators,
        \item multi-GPU features for high performance computing of large tomographs,
        \item automatic data loaders,
        \item pre-processing steps for proton CT;
    \end{itemize}
    \item PyTorch wrapper implementation (given the growing interest of data-driven methods in reconstruction).

\end{itemize}


\subsection{Tomographic reconstruction and iterative algorithms}\label{sec:tomo_iter}



The CT measuring (or forward) process can be modelled mathematically using the Radon transform, which is an integral operator that considers line integrals over the domain. The most common approach to CT reconstructions is to consider the large scale linear system of equations arising from the discretization of such measuring process, which is of the form:
\begin{equation}\label{eq:linear}
    Ax+\tilde{e}=b,
\end{equation}
where $A$ is the (known) system matrix describing the forward operation; $x$ is the (unknown) vector of image values in lexicographical order; $b$ is the (known) vector of measured (noisy) linear attenuation values in lexicographical order; and $\tilde{e}$ represent the noise affecting the measurements (both measurement noise and model errors introduced by the linearization of the real forward operator).  The least-squares solution to equation~\eqref{eq:linear} can be found using the Moore–Penrose pseudoinverse of the forward operator:
\begin{equation}\label{eq:pseudo}
    x=A^\dagger b,
\end{equation}
however, this can be far from the true solution $x$ due to noise amplification.

For CT problems, approximations of the analytical solution to the original Radon transform problem exist, the discretization of which is also known as the Filtered Backprojection (FBP) in the CT community, or the Feldkamp-Davis-Kress (FDK) algorithm for Cone Beam CT (CBCT). 
These methods produce useful reconstructions when the problem is not very ill-posed and there is a low level of noise in the measurements. This is also the most used approach in practice, fostering the wrong belief that high quality, high X-ray dose and over-sampled data are required to obtain meaningful CT reconstructions. 

However, there are great advantages associated to reducing the X-ray dose and scanning times in terms of cost and safety. Moreover, it is sometimes the case where the physical limitations of the problem only allow a limited sampling space (e.g. limited angle tomography). In all these scenarios, the corresponding problem in equation \eqref{eq:linear} is ill-posed, or, in other words, small perturbations on the measurement data can cause large perturbations in the reconstruction. This means that, in the presence of noise, the approximate solution obtained using equation \eqref{eq:pseudo} will be corrupted or even totally dominated by the noise amplification. To alleviate this, one needs to resort to regularization. One of the most well known approaches to regularization is the use of variational methods, which consist on solving the minimization problems of the form:
\begin{equation}\label{eq:min}
    \min_x D(Ax,b)+\alpha R(x),
\end{equation}
given a distance function $D(\cdot,\cdot)$ and a regularization function $R(\cdot)$ that penalizes constraints in the solution and can be related to the \textit{prior} of the image if one sets the original problem \eqref{eq:linear} in the Bayesian framework \cite{Calvetti2007Bayesian}. Here, the \textit{hyperparameter} $\alpha$ controls the trade-off between the image fitting the measured data and the image fitting the prior information. Usually, the choice of the distance function $D(\cdot,\cdot)$ is informed by the assumptions on the distribution of the unknown noise. In CT applications, the most common choice for this is to use the 2-norm, because the measured noise is well approximated by a Gaussian distribution after the standard log-transform of the data \cite[Chapter
2.3.2]{doi:10.1137/1.9781611972344}. However, some approaches consider the noise associated to the photon counts in the raw data, which has a Poisson distribution, so a different distance function $D(\cdot,\cdot)$, called Kullback–Leibler (KL) divergence, needs to be considered instead. 

Different algorithms for given combinations and choices of $D$ and $R$ have been provided over the years by the optimization and linear algebra communities, with different behaviours and convergence properties that can lead to superior image quality. These methods, almost always iterative in nature, are covered by the generic term \textit{iterative algorithms}. For example, a common method to solve any convex optimization method is the Gradient Descend (GD) method. Given the distance function of the 2-norm and no regularization, one can obtain a solution of equation \ref{eq:min} using the iterative update
\begin{equation}\label{eq:gradient}
    x^{k}=x^{k-1}+\lambda A^T(Ax^{k-1}-b),
\end{equation}
where $k$ is the iteration number and $\lambda$ a step size, desirably inversely proportional to the norm of $A$. Similarly, a common algorithm to solve the same problem given the KL-divergence is the Maximum Likelihood Expectation Maximization (MLEM) algorithm, with an update of the form
\begin{equation}\label{eq:mlem}
    x^{k}=\frac{x^{k-1}}{A\mathrm{1}}A^T\frac{b}{Ax^{k-1}},
\end{equation}
given $\mathrm{1}$ a the all-ones vector. 

There is an abundance of alternative methods to solve CT reconstruction problems, some accepting variants with explicit regularization and some with implicit regularization properties instead. These have significant differences in convergence behaviour and solution characteristics. TIGRE provides a platform with a vast array of the most common solvers, implemented in such a way that they can be directly used with real datasets. In particular, TIGRE has the following algorithms, categorized in different classes: 
\begin{itemize}
    \item \textbf{Direct methods}. These methods implement an analytic expression of the pseudoinverse, as in equation \eqref{eq:pseudo}, or suitable approximations of it. Included are FBP and FDK\cite{fdk}, with various filter types and geometric correction methods, such as Wang~\cite{wang2002x} and Parker~\cite{wesarg2002parker} weights for detector displacement and angular range corrections, respectively.
    \item  \textbf{Kaczmarz-type algorithms}. These methods are generalizations of GD: the approximate images are updated similarly to \eqref{eq:gradient} but with a reweighing of the rows and columns to normalize for length. They were rediscovered as the Algebraic Reconstruction Technique (ART)~\cite{ART} in 1970 for CT with the inclusion of non-negativity constraints. Different modifications include the Simultaneous version (SART)~\cite{SART}, the Ordered Subsets version\footnote{See Section \ref{sec:OS}} (OS-SART)~\cite{OS_SART} and the confusingly named Simultaneous Iterative Reconstruction Algorithm (SIRT)~\cite{SIRT}. The main difference between them is the update size. 
    \item \textbf{Projection Onto Convex Sets (POCS) algorithms}. These algorithms allow solving two convex optimization problems jointly, in the TIGRE implementation, these correspond specifically to the ones related to the data constraints and the regularization. Given a 2-norm data constraint, and various flavours of Total Variation (TV) regularization (promoting pairwise smooth images), several POCS-like algorithms are available. These include the well-known Adaptive Steepest Descend version (ASD-POCS)~\cite{ASD_POCS} and the Projection-Controlled Steepest Descent (PCSD)~\cite{liu2015reconstruction}, which reduces the required parameter selection. Versions of these two algorithms exists for the Adaptive weighed TV norm~\cite{liu2012adaptive,lohvithee2017parameter}\footnote{See Section \ref{sec:features}}, and OS versions exist for the first with both TV norms. A version that uses a Bregman outer iteration is also available, B-ASD-POCS-$\beta$~\cite{B_ASD_POCS_beta}.
    \item \textbf{Krylov Subspace algorithms}. These are methods that project the original optimization problem into small relevant subspaces of increasing dimension, producing fast converging algorithms with intrinsic regularization properties, that can be used in combination with explicit 2-norm regularization. In TIGRE, the following are implemented allowing, when possible, to compute the regularization parameter automatically: CGLS~\cite{hestenes1952methods}, LSQR~\cite{paige1982algorithm}, hybrid LSQR~\cite{paige1982algorithm}, AB/BA-GMRES~\cite{hansen2022gmres}, LSMR~\cite{fong2011lsmr}. Moreover, two different methods that allow for approximations of the TV regularization are also implemented: IRN-TV-CGLS~\cite{wohlberg2007iteratively} and hybrid-fLSQR-TV~\cite{gazzola2019flexible}. 
    \item \textbf{Statistical minimization algorithms}. These methods are based on the understanding of the problem \eqref{eq:min} as a Bayesian minimization problem tied with the KL-divergence distance function. MLEM has already been introduced in equation \eqref{eq:mlem}, and its OS-counterpart, widely used in emission tomography, OSEM\cite{hudson1994accelerated}, is also available. 
    \item \textbf{Proximal algorithms}. More recently, optimization algorithms that combine \textit{proximal} operators have been shown to be successful in image reconstruction. For the 2-norm and TV proximal operators, TIGRE provides the Fast Iterative Shrinkage-Thresholding Algorithm (FISTA)~\cite{beck2009fast} and further improvements for faster speed~\cite{liang2018faster}. Moreover, TIGRE also includes the SART-TV algorithm~\cite{SART_TV}, which was not originally proposed within this framework, but that can also be interpreted as a proximal method.
\end{itemize}

Some of the algorithms mentioned above were already available in the original TIGRE release, but many have been added over the years into its current version. New introductions to TIGRE are described in more detail in the following sections wherever appropriate, but the authors suggest reading the original articles for a more in depth description. A summary of all the available algorithms (as of this date), can be found in \ref{Appendix1}, along with their corresponding citations, both in the original literature and in the corresponding TIGRE-specific implementation papers.


\subsection{Improvements on the iterative reconstruction tools in TIGRE}
This section describes the enhancement of iterative reconstruction algorithms available it TIGRE since the first release. This section is divided into three, describing first the addition of Krylov Subspace algorithms to TIGRE, then Ordered Subsets (OS) types of block-action algorithms, and finally a brief description of various minor yet useful features and regularization tools included since its first release. 

\subsubsection{Krylov subspace algorithms}\label{sec:krylov}
Krylov subspace methods are a class of iterative solvers for linear equations of the form \eqref{eq:linear}. Because they are naturally regularizing, they are especially suited for inverse problems. Moreover, these are projection methods which only require matrix-vector products with the system matrix (and its adjoint), so they are very suited to solve large-scale problems. The TIGRE toolbox features a collection of some of the most standard Krylov Subspace methods for non-square matrices (CGLS, LSQR, LSMR), possibly
in combination with Tikhonov regularization, as well as recent developments including total
variation (TV) regularization and mismatched backprojectors. An in detail explanation of the algorithms as well as its TIGRE implementation can be found in \cite{KrylovTigre2023}.

\subsubsection{Ordered Subsets algorithms}\label{sec:OS}
Ordered Subsets (OS) methods are named after the fact that only a subset of the data is used to update 
the solution of the tomographic reconstruction problem at each iteration. Particularly, at iteration $k$, the solution $x^k$ is updated using a row-wise subsample of the matrix $A$, namely $A_s$, and therefore only the corresponding part of the measurements $b$, namely $b^s$. These algorithms are most commonly known as \textit{block-iterative methods} in the mathematical community~\cite{censor1997parallel}. For example, similarly to equation \eqref{eq:gradient}, an approximate gradient step for the 2-norm distance for solving equation \eqref{eq:min} with OS would be
\begin{equation}
    x^k = x^{k-1}+\lambda A^T_s (A_sx^{k-1}-b_s),
\end{equation}
where $s$ defines a row-wise block of the matrix $A$ which changes at each iteration.

Compared to algorithms that use the entire matrix in their update step (e.g. SIRT), OS algorithms require more (partial) updates to use all the measured data. However, they are well known for accelerating the rate of convergence, in the sense that many fewer iterations (defined as the number of times that the whole matrix $A$ is seen) are required to  produce a reconstruction of comparable quality.  On the other extreme, algorithms that update the solution projection by projection display faster converge in terms of the number times that one needs to access the whole system matrix, but they are often very slow in wall-time due to the large computational cost of having to update the solution a higher amount of times. Therefore, OS algorithms produce a reasonable middle ground algorithm between fast mathematical convergence and fast runtime~\cite{wang2004ordered}. This is why their use in CT is common, particularly in Kaczmarz-like methods (e.g., OS-SART~\cite{OSSART}) and the most used algorithm in emission tomography like PET and SPECT is of this family (namely OS-EM~\cite{hudson1994accelerated}). Moreover, there are many algorithms that can be used in combination with an OS scheme. This is the case of methods designed for solving problems with explicit regularization that require update steps that are only related to the data-minimization term, without any particular prescriptions on the form of this update. 
In the original version of TIGRE, OS-SART was available. Since, ASD-POCS, AwASD-POCS, PCSD, AwPCSD and MLEM now have an OS counterpart.


When invoking these algorithms, a parameter that regulates the size of each block (in number of projections per block) can be selected by the user, as seen in Snippet \ref{code:OS} for ASD-POCS and its block-action counterpart. Updating can be done sequentially, randomly or via maximum information updates. 


\begin{minipage}{\linewidth} \begin{lstlisting}[language=iPython,caption=OS-ASD-POCS algorithm in Python,label=code:OS]
import tigre.algorithms as algs
n_iter=10
res_asd_pocs = algs.asd_pocs(projections, geo, angles, n_iter)
res_os_asd_pocs = algs.os_asd_pocs(projections, geo, angles, n_iter, blocksize=20)
\end{lstlisting} 
\end{minipage}

\subsubsection{Algorithm features}\label{sec:features}
New features that are useful for most algorithms have been introduced to TIGRE since its original release. Here is a short summary:
\begin{itemize}
    \item \textbf{Residual computing}. The residual norm history $\|Ax^k-b\|$, for each iteration number $k$, is an interesting value to track when comparing iterative methods, as it directly describes how well the current reconstruction of the image fits the data. This computation per iteration is costly (it requires an extra forward projection) but TIGRE now allows users to ask for this residual, if desired, for algorithm performance comparison. Note that Krylov methods include a way of estimating this quantity at no extra cost per iteration. Last, note that an exact fit is not desired, as that would also fit the noise present within the sinogram $b$.  
    \item \textbf{Ground truth comparison}. The error norm histories $\|x_{gt}-x^k\|$, given a known ground truth image $x_{gt}$, are also of capital interest when studying algorithm performance. Note that, due to the ill-posedness of the problem, metrics solely based on the residual norm might be misleading. TIGRE now allows to input a ground truth to be compared against the current iterate during reconstruction. 
    \item \textbf{Automatic regularization selection}. Krylov subspace algorithms have a single regularization hyperparameter when used in combination with variational regularization, i.e. $\alpha$ in equation \eqref{eq:min}, and there exist extensive theory on how to chose this parameter optimally~\cite{Chung2024Hybrid}. Some algorithms in TIGRE include therefore automatic regularization parameter selection. If the noise-level value is either known or can be estimated, it can be inputted as a parameter of the algorithm to guide the choice of regularization parameter in the optimization problem, using the so-called Discrepancy Principe (DP)~\cite{Morozov1966DP}. Otherwise, the Generalized Cross Validation (GCV)~\cite{Golub1979GCV} is used instead to update the regularization parameter at each iteration. For more details on the implementation see~\cite{KrylovTigre2023}.
    \item \textbf{Adaptive weighted total variation}. The well known Total Variation (TV) regularization promotes piece-wise constant reconstructions. However, this can sometimes have a significant effect in the shape of sharp edges, often the most important part of CT image analysis. Adaptive weighted TV (AwTV)~\cite{liu2012adaptive}, introduces a penalty term $w_{i,j}$ in order to only account for small variations in pixel values (mostly caused by noise), but not penalize large ones (which are more likely to represent real edges). Thus, the new TV term with adaptive  weights (for the two-dimensional case) is:

    \begin{equation}
    \mathrm{AwTV}(\mathbf{x}) = \sum_{i,j=1}^{n-1}\sqrt{w_j\left(x_{i,j} - x_{i,j-1}\right)^2 + w_i\left(x_{i,j} - x_{i-1,j}\right)^2},
   \end{equation}
    with weights defined as
    \begin{equation}
        w_{i} = e^{-\left(\frac{x_{i,j} - x_{i-1,j}}{\delta}\right)^2}, \qquad w_{j} = e^{-\left(\frac{x_{i,j} - x_{i,j-1}}{\delta}\right)^2},
    \end{equation}
    with a parameter $\delta$, that produces exactly the TV norm when $\delta \rightarrow \infty$.
    The AwTV norm can be now used with any POCS-type algorithm, and a gradient descend minimizer of the function is available for users. 
    \item \textbf{Data redundancy weighting on iterative algorithms}. The data redundancy weighting applied in FDK for the cases where the detector is offseted have been shown to improve reconstruction of SART and gradient descend-like algorithms~\cite{sanctorum2021extended,bian2012optimization}. All algorithms that accept this type of weighing now have it by default. 
\end{itemize}

\subsection{Python implementation}
On its original release, TIGRE was only a MATLAB toolbox, which limited its definition as a free and open source toolbox, as MATLAB is neither free nor open source itself. Now, all TIGRE's MATLAB codebase has been duplicated into the Python language. 
The structure and functionality of both implementations is the same, however, few technical differences exist: such as memory unrolling ordering for multidimensional arrays, the use of Cython rather than MATLABs MEX interface or the import hierarchy that does not exist in its counterpart. 

Moreover, there is a notable difference between the MATLAB and the Python releases which goes past the technical aspects: the Python version uses an Object Oriented Programming (OOP) paradigm with a system of classes and inheritance, where all algorithm are children of a 
\lstinline[language=iPython]{IterativeReconAlg}
class that sets some common methods and attributes and offers a framework for other algorithms to build upon. Examples of the Python interface are given across this document. 

\subsection{Geometric flexibility}
In its original form, TIGRE allowed for most standard circular CBCT geometries, such as detector shifts, or reconstructions of images not centered in the axis of rotation. However, since the first publication \cite{TIGRE}, it has been evident that more complex geometries are both required for several non-standard but common scans, such as helical CT, tomosynthesis~\cite{vedantham2015digital} or laminography~\cite{o2016recent}, among others. Moreover, researchers on CT methods are showing an increased interest in exploring with complex scanning paths and setups, and having the right computational tools for both producing simulations and reconstructions can be of crucial importance. Thus, TIGRE's geometric flexibility has been updated with several important features:
\begin{itemize}
    \item \textbf{Detector in-place rotation}. The detector is assumed to lie in the orthonormal plane to the source-to-detector line. However, this is not necessarily true either by design or by mechanical inaccuracies. For example, scanners settings like those used in laminography or tomosynthesis purposely have measurements where the detector is not orthonormal to the source. This can be easily expressed (equivalently) by a rotation of the detector instead of a circular CBCT trajectory. Moreover, many scanners have mechanical inaccuracies that arise after use, and calibration methods exists to correct such changes algorithmically~\cite{yang2006geometric}. TIGRE now accepts in-place $(\phi,\theta,\psi)$ rotation angles for the detector. 
    \item \textbf{Center of Rotation (CoR) correction}. In \textmu-CT scans, where the sample rotates, rather than the machine, it is standard that the axis of rotation is slightly shifted away from the source-to-detector line. While this could be described with a changing image combining a detector shift and a detector in-place rotation, it is significantly easier to describe as a shift in the axis of rotation itself, as TIGRE allows now. 
    \item \textbf{Arbitrary axis of rotation}. To fully describe any geometry, one would also need to allow for any arbitrary axis of rotation, not only a vertical one, as a standard circular CBCT machine produces. This enables the use of TIGRE not only on experimental lab setups, but also in more standard C-arm CBCT machines. The implementation allows now to include Euler ZYZ $(z_1,y_1,z_2)$ angles as gantry angle rotations. 
    \item \textbf{Curved detectors}. While most CBCT detectors are planar, 2D CT detectors are often curved (with equidistant source-detector distance in each pixel). Recently, some experimentation is being done on curved CBCT detectors. In that spirit, TIGRE now allows to consider curved detectors by applying a ``flattening'' transformation. 
    \item \textbf{Per-projection geometry}. In order to allow a fully flexible geometry for any arbitrary CBCT scan that exists and could be imagined, the defined geometry parameters should be able to vary for each projection measured. Thus, TIGRE now allows for all geometry defining parameters, except the image and detector size in both voxels and real world units, to be defined individually for each projection. This allows for arbitrary scanning geometries, as shown in~\cite{hatamikia2021toward}. 
\end{itemize}

\begin{figure}[h]
    \centering
    \includegraphics[width=0.75\linewidth]{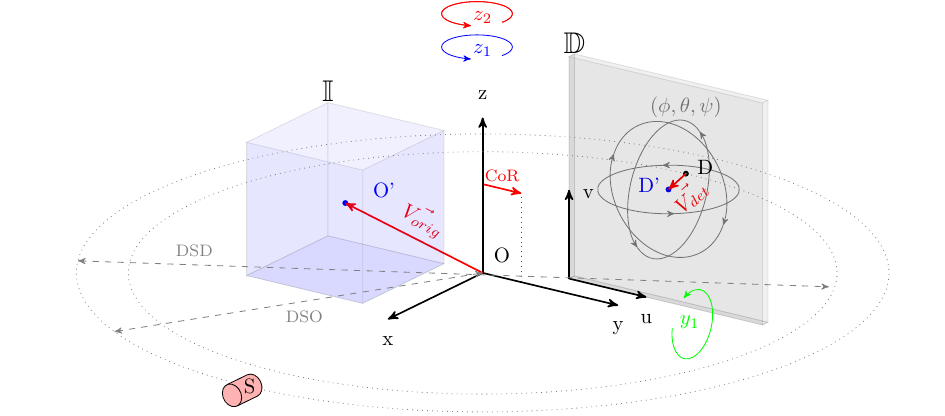}
    \caption{Diagram of geometric parameters of TIGRE for a single X-ray projection.}
    \label{fig:geo}
\end{figure}

All this considered, the geometric diagram of TIGRE shown in the original release (Figure 1 in the original article~\cite{TIGRE}), should be updated to Figure \ref{fig:geo} presented here. 

Figure \ref{fig:geo} can be better understood if we provide a brief description of the default scanning geometry assumed by TIGRE. Here, the scanner is assumed to rotate around a fixed, static object. This volume is centred at the Oxyz axis, which does not rotate with the machine, but the origin O is allowed for offsets, so some other scanning modes, such as the helical scanning can be realized.. The detector is assumed to be centred in the source-origin ($\text{S}$-$\text{O}$) line, at point $\text{D}$ (with distance DSO from the origin). An auxiliary axis exists for the detector, uv, with an origin in the bottom left corner of the detector and always aligned with the detector itself. The rest of the parameters in the geometry definition of TIGRE, which allow for modifications of the default settings, are described in \ref{Appendix_geo}.

\subsection{Multi-GPU support}
Since CT reconstruction is a very computationally heavy task, it benefits from access to any computational resources available. With the raise of multi-GPU machines, now a staple computational resource in most CT laboratories and hospitals alike, TIGRE has been extended to support multiple-GPU computing~\cite{biguri2020arbitrarily}. This allows faster computations in multiple GPU machines and also enables larger-than-memory computations. 

The TIGRE implementation of this support is done via a novel memory management approach that seamlessly distributes any computationally heavy operation (forward and backprojector, and regularization optimizers) across multi-GPUs (either user selected, or all available), in such a way that the computation scales linearly with the number of GPUs. In addition to providing a faster execution time, the splitting algorithm that TIGRE implements can be used to reconstruct images that are larger than the GPU memory. This is, for example, a very common scenario in \textmu-CT applications, where both the projection and the image sizes are often  bigger than most commonplace GPU DRAM sizes. This allows, as long as the CPU RAM is large enough, to reconstruct large images on arbitrarily small GPUs with a minuscule extra computational footprint added. This is achieved by optimized overlapping of memory copies and computational kernels in the GPUs, such that memory is only copied in or out of the GPU while computing is performed, and the projectors almost never need to wait for a memory transfer to finish. More information can be found in \cite{biguri2020arbitrarily}. Moreover, in the cases where the CPU RAM is still too small for the problems in hand, TIGRE also allows the use of swap memory. However, this will be slower for two main reasons: the novel memory management system cannot be used in this case, and swap memory is slower than RAM.  Lastly, the TIGRE toolbox includes a GPU-management system that allows the user to select which GPUs they want to use, either by name or GPU id. This is an essential tool for users computing in heterogeneous platforms or shared resources.  Code Snippet \ref{code:GPU} showcases how this can be selected, and passed into an algorithm. 

\begin{minipage}{\linewidth}
\begin{lstlisting}[language=iPython,caption=Optional GPU selection in Python,label=code:GPU]
from tigre.utilities import gpu
gpuids = gpu.getGpuIds("GeForce RTX 2080 Ti") # Will only select these GPUs 

gpuids = gpu.GpuIds()               # Get all GPUs
gpuids.devices = int32(2:3)         # Only use IDs 2 and 3

import tigre.algorithms as algs
# Define projections, geo, angles, n_iter
res_cgls = algs.cgls(projections, geo, angles, n_iter,gpuids=gpuids)
\end{lstlisting} 
\end{minipage}

It is worth noting that having this multi-GPU features is crucial for the \textmu-CT community to gain access to the iterative reconstruction field, allowing the testing of large acquisitions with modern reconstruction methods~\cite{SUN2023102852,tekseth20244d,du2018x}.
\subsection{Data Loaders}
A key and often non-trivial feature of a user-driven toolbox is the capability of loading datasets directly from the output of the devices that acquired the measurements. This is particularly true for CT, because a standard file format for sinograms does not exist. On the contrary, each scanner manufacturer has a unique way of storing the measurements and scanning parameters, often changing between different versions of software of the same manufacturer. Correct data and system geometry loading is often key. 
In TIGRE, data loaders for 6 different major CT manufacturers are available, 2 from the medical community (Philips, Varian) and 4 from the \textmu CT community (Comet Yxlon, Nikon, Bruker, Diondo). Additionally, the Data Exchange (DXChange) format~\cite{de2014scientific}, proposed for synchrotron tomography sharing, is also supported. Among these, the Varian loader contains a much higher set of functionalities for the preprocessing of the data, and TIGRE emulates the Varian CBCT machines pipeline\footnote{The authors do not have proprietary information on Varian software, the functions here are reverse engineered from domain knowledge, not private information.}. This is explained in more detail in the article by Yi \textit{et al.}~\cite{VARIAN}.

The data loaders are included in TIGRE such that loading any complex dataset can be called in a single line of code containing image loading, geometry reading and any other preprocessing operations needed, such as applying the Beer-law to the projection. The following  snippet highlights how it is used:

\begin{minipage}{\linewidth} 
\begin{lstlisting}[language=iPython,caption=Data loading script in Python,label=code:data_loader]
import tigre.utilities.io as tigreio
projections, geometry, angles = tigreio.NikonDataLoader('/path/to/dataset')
\end{lstlisting} 
\end{minipage}

Few other features are available, such as the ability to load partial datasets, or collections of datasets, when available.

\subsection{Proton Computed Tomography}
Since the early 2000s, proton therapy centers have been playing an increasingly important role in cancer therapy. This is because the unique properties of these particles allow treating deep-seated tumors with high precision (sharp dose-deposition). In order to fully exploit the potential of proton therapy, the planning CT could be measured with protons as well. This would allow bypassing the inaccuracies in treatment planning that arise from the conversion of Hounsfield units (obtained from a standard planning CT) to relative stopping powers of protons, and could effectively improve therapy \cite{Meyer_2019}. 

The properties of an imaging setup for proton computed tomography (pCT), which was first described in \cite{schulte2004conceptual}, differ from those of standard CT, and hence both the modelling and the subsequent reconstruction algorithms need to be modified accordingly. The main difference is that proton paths through matter cannot be accurately approximated by a straight line, due to multiple Coulomb scattering, and therefore information obtained from particle tracking has to be included in the reconstruction process. TIGRE includes a novel preprocessing step to do this \cite{kaser2022extension}, which allows generating optimized (rebinned) radiographs from the pCT measurements. This method is based in \cite{collins2016maximum}, where a maximum-likelihood approach combined with a cubic spline estimate for the proton path was used to obtain proton radiographs with high resolution. For improved pixel weighting in the radiographs, a cylindrical object hull can be used as input in the newly implemented binning step in TIGRE. While the binning itself is implemented in CUDA, the user can call the method, after defining the geometry, by a one-line MATLAB command:

\begin{minipage}{\linewidth} \begin{lstlisting}[language=Matlab,caption=Binning of pCT data into optimized radiographs,label=code:prad]
geo = % Define geometry as usual
% New geonetry parameters
geo.DSID = 300; % Distance between source and upstream detector.
geo.DSD = 700; % Distance between source and downstream detector.
geo.hull = [150; 150; 0; 40]; % Convex hull 

pRad = pCTCubicSpline_mex(data.posIn, data.posOut, data.dirIn, ...
    data.dirOut, data.Wepl, eIn, geo);
\end{lstlisting} 
\end{minipage}

In the code snippet, \textit{pRad} refers to one optimized proton radiograph: for a full pCT, this step has to be repeated for each radiograph. The data input is a list of the proton positions and directions upstream and downstream the patient, the water-equivalent path length measured for each proton, the incident proton energy and the pCT imaging geometry (which now also contains the locations of the particle tracking detectors and optionally an object hull). A demo file has been added to TIGRE to facilitate easier usage. After all proton radiographs have been preprocessed, any implemented reconstruction algorithm in TIGRE can be used seamlessly without any further adaption for pCT.

\subsection{PyTorch operator wrappers}

State-of-the-art CT reconstruction includes data-driven methods, many of which use CT operators (forward and backprojection) within the model, e.g. Learned Primal Dual~\cite{adler2018learned}. As for the time of writing, Torch, and in particular PyTorch~\cite{NEURIPS2019_bdbca288}, is the main machine learning library used in academia and industry. TIGRE now includes an optional PyTorch binding that allows TIGRE projectors to be included inside PyTorch models, and be treated as a linear differentiable operator by the automatic differentiation engine. Code Snippet \ref{code:torch} shows how TIGRE auto-differentiable operators can be created to use within PyTorch. 

\begin{minipage}{\linewidth}
\begin{lstlisting}[language=iPython,caption=Use of TIGRE's PyTorch bindings,label=code:torch]
from tigre.utilities.PyTorch_bindings import create_PyTorch_operator

geo = tigre.geometry(mode="fan")
angles = np.linspace(0, np.pi, 200)
ax, atb = create_PyTorch_operator(geo, angles)
# Now ax and atb can be used inside autograd-enabled pipelines
import torch
device = torch.device("cuda:1")
input_volume = torch.randn([2,geo.nVoxel[0], geo.nVoxel[1],geo.nVoxel[2]],
                requires_grad=True).to(device)

sino_batch = ax(input_volume) # returns batched torch tensor
image_batch = atb(sino_batch) 
\end{lstlisting}
\end{minipage}



While this allows for TIGRE to be used within PyTorch, it is important to mention that the embedding of the operators is not as sophisticated as it is with other tools, e.g. tomosipo~\cite{tomosipo}. Particularly, TIGRE's approach requires the memory to be transferred from a PyTorch tensor to a NumPy array, which more GPU$\leftrightarrow$CPU memory transfers are needed than in tomosipo.

\section{Results} \label{Sec:results}
In order to showcase the usability of TIGRE in a variety of real datasets, we present several experiments alongside the code needed to reproduce them. The first example concerns a clinical scenario with patient data. The second example consists of a synchrotron parallel beam tomographic reconstruction. The third example corresponds to a Monte Carlo simulation of a proton CT problem (due to its early stages of development as a device). Next, a neutron CT scan is showcased. Finally, the last example involves an industrial \textmu CT dataset. 

It is important to note that this documents is not making scientific statements about which reconstruction techniques are best in any of the experiments presented here. This is because, often, the performance of CT imaging reconstructions is heavily dependent on secondary tasks, e.g. the best image for segmentation is not necessarily the best image for diagnosis. Similarly, human eyes are biased towards natural looking images, rather than the maps of scalar fields outputted in CT reconstructions. Moreover, experts from different fields have sometimes trained their perception to filter out noise, e.g. a radiologist knows how to identify the differences between an artifact and a lesion in medical CT. Therefore, the only claim we make in this work is that reconstructions obtained using different algorithms are different, and that they may be of interest in different scenarios. For this reason, there is a real value on making more reconstruction algorithms easily available for both clinician and algorithm developers, as well as platforms to promote knowledge transfer between them. 

It is also worth noting that all the experiments presented in this section have been run in a personal high-end desktop (Ryzen 5, 32GB RAM, RTX 4070), and not in a large workstation, highlighting the power of TIGRE to produce computationally fast results in day-to-day use machines.

\subsection{Medical dataset from a Varian CBCT machine}
With the data loading and pre-processing tools added to TIGRE, particularly for the Varian range of onboard CBCT machines, iterative algorithms can easily be utilized in clinical images. This experiment shows how to read and reconstruct such images, in particular from a Varian Edge CBCT situated at Beijing Cancer Hospital. 

\begin{figure}[h]
    \centering
    \includegraphics[width=0.75\linewidth]{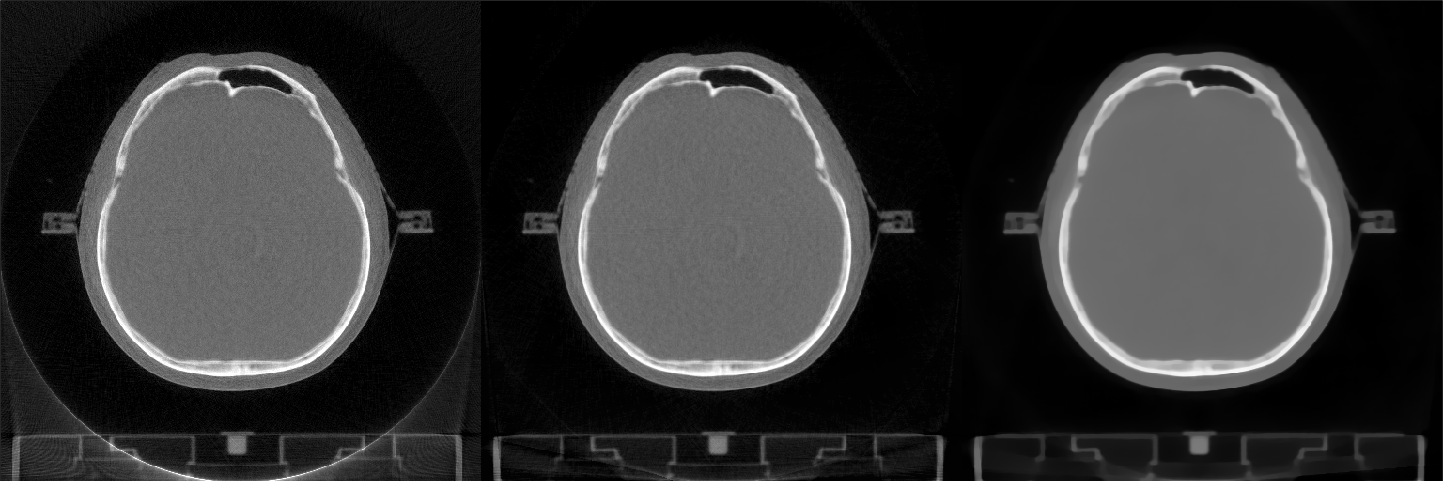}
    \includegraphics[width=0.75\linewidth]{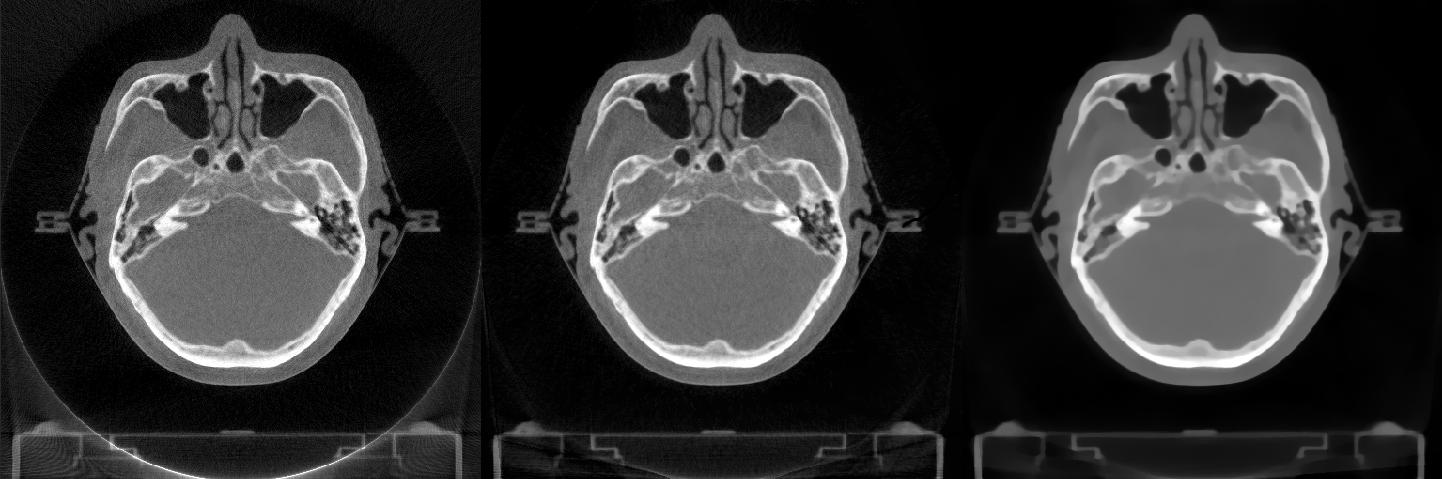}

    \caption{Reconstruction from a Varian Edge machine, 200$^{\circ}$ arc head scan at Beijing Cancer Hospital. Rows show different slices of the head, columns show the FDK, OS-SART (50 iterations) and OS-ASD-POCS (50 iterations) reconstructions, from left to right. Visualization is linear attenuation coefficients in range [0-0.05].}
    \label{fig:head_varian}
\end{figure}

The first dataset is a head scan with projections of size $1024\times768$ taken on a limited arc range of 200$^{\circ}$, uniformly sampled with 493 projections, where the size of the reconstruction is $512\times512\times362$ voxels.  The projections are loaded and pre-processed with the data loader, and then reconstructed using three different algorithms: FDK, OS-SART, and OS-ASD-POCS. The latter includes TV regularization to reduce the noise in the reconstruction. The script that produces the reconstructions can be seen in Snippet \ref{code:Varian_experiments} and two different slices of the head at different craneo-caudal slice values can be seen in Figure \ref{fig:head_varian}. Each reconstruction took less than 5 minutes (few seconds for FDK) in a personal desktop setup. In the reconstruction, one can see that OS-SART and OS-ASD-POCS produce images with less pixel-level noise than FDK, particularly noticeable in OS-ASD-POCS, as it has TV regularization, which promotes flatter images.

\begin{minipage}{\linewidth} 
\begin{lstlisting}[language=Matlab,caption=MATLAB code to reconstruct Varian CBCT data,label=code:Varian_experiments]
% Load TIGRE
InitTIGRE;
% Load
[projections, geo, angles] = VarianDataLoader("./path/to/Head/scan");
% Reconstruct
recon_fdk     = FDK(projections,geo,angles);
recon_os_sart = OS_SART(projections,geo,angles,50);
recon_tv      = OS_ASD_POCS(projections,geo,angles,50);
\end{lstlisting} 
\end{minipage}

\subsection{Synchrotron dataset}

Synchrotrons are large electron accelerators that have become an essential experimental facility for scientific discovery because they produce high quality X-rays of monochromatic energy and low noise statistics. This can be used to obtain extremely high quality CT images, and it also allows for also more complex experiments, such as X-ray crystallography, X-ray diffraction imaging and several different modalities. For this reason, they are routinely used for material science, biology and many other applications. 

\begin{figure}
    \centering
    \includegraphics[width=0.75\linewidth]{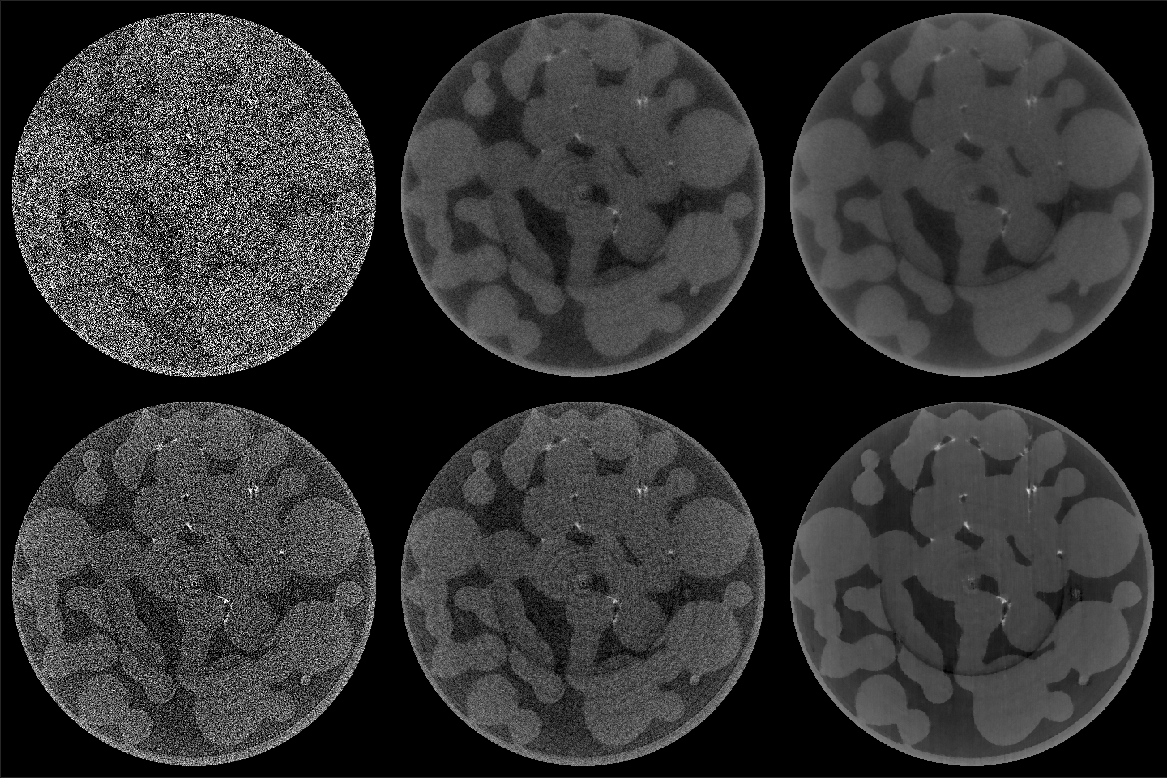}
    \caption{Synchrotron dataset acquired with settings for a dynamic in-situ experiment. Reconstructions are, from left to right and top to bottom: FBP with linear filter, LSQR (20 iterations), FISTA (80 iterations), FBP with Hamming filter, SIRT (150 iterations) and OS-ASD-POCS (20 iterations).}
    \label{fig:syncro}
\end{figure}

While synchrotron experiments often produce very high quality data, and thus iterative reconstruction is often unnecessary, this is not always the case. In this experiment, taken from the TomoBank data repository\cite{de2018tomobank}, a single time-frame from a dynamic acquisition is reconstructed. Often, scientist want to perform in-situ experiments and image the evolution of certain processes. In these cases, the acquisition time has to be severely reduced with respect to the optimal time one would need to  capture the dynamics of the experiment. Thus, imaging data is very low dose and therefore contains very high levels of noise, which produces severe artifacts in the commonly used FBP reconstruction, as seen in the top-left of Figure \ref{fig:syncro}. In this experiment, the projections are of size $960\times900$ and sampled over 900 angles, uniformly distributed around 180$^{\circ}$ range. The reconstructed images have $960\times960\times 900$ voxels. 

In this experiment, several algorithms are used to reconstruct the same data to showcase the different nature of the resulting images. In Snippet \ref{code:syncro} we can see the code that produces the images in Figure \ref{fig:syncro}, where six different reconstruction can be seen. The leftmost column shows FBP with the standard ramp filter on top and FBP with a noise-rejecting Hamming filter on the bottom. The second column shows two iterative reconstruction algorithms with no explicit regularization function, LSQR on the top and SIRT on the bottom. The last column shows TV-regularized algorithms, namely FISTA and OS-ASD-POCS, with their higher noise-rejection behavior. 

\begin{minipage}{\linewidth} 
\begin{lstlisting}[language=Matlab,caption=MATLAB code to reconstruct synchrotron data,label=code:syncro]
% Load TIGRE
InitTIGRE;
% Load
[projections, geo, angles] = DXChangeDataLoader("./data_file.h5");
% Reconstruct
recon_fbp         = FBP(data, geo, angles);
recon_fbp_hamming = FBP(data, geo, angles,'filter','hamming');
recon_lsqr        = LSQR(data,geo,angles,20);
recon_sirt        = SIRT(data,geo,angles,150);
recon_fista       = FISTA(data,geo,angles,50,'tviter',50);
recon_os_asd_pocs = OS_ASD_POCS(data,geo,angles,20,'blocksize',80);
\end{lstlisting} 
\end{minipage}

\FloatBarrier
\subsection{Proton Computed Tomography}

Using the code demonstrated in Code~Snippet~\ref{code:prad}, optimized proton radiographs were calculated in \cite{kaser2022extension}, using as input parameters the measured water-equivalent thickness, the initial proton energy, the upstream and downstream positions and directions for each proton and the pCT geometry. The pCT geometry closely follows the standard geometry definition in TIGRE, however, the locations of upstream and downstream detectors (as opposed to only one detector in conventional CT), as well as the convex hull (here implemented as a cylinder) have to be declared.

For this example, Catphan phantoms modules  \cite{ctp_hr} (15~cm diameter) were simulated with the Monte Carlo toolkit Geant4 \cite{Agostinelli2002hh} and irradiated with 225~protons/mm$^2$. The synthetic measurements were then combined into optimized binned radiographs in TIGRE with the new pCT implementation, to create 90 radiographs (4 degree steps over a range of 360 degree) of the high resolution phantom (PMMA body with aluminium line pair insets).  The experiment data was obtained from the the ``non-ideal'' parallel simulation case (realistic energy resolution of $\Delta E / E = 1\%$, detector resolutions of $\sigma_s$ = 0.15~mm and detector thicknesses of 300~$\mu$m), that was described in full detail in \cite{kaser2022extension}. These optimized radiographs can be used as CT-like data in TIGRE and hence inputted to the implemented algorithms without further adaptation.

\begin{minipage}{\linewidth} 
\begin{lstlisting}[language=Matlab,caption=Reconstruction of optimized proton radiographs.,label=code:pct]
% Load optimized proton radiographs 
recon_fdk = FDK(projections,geo,angles);
recon_os_sart = OS_SART(projections,geo,angles,100);
recon_tv = ASD_POCS(projections,geo,angles,20);
\end{lstlisting} 
\end{minipage}

\begin{figure}
    \centering
    \includegraphics[width=0.32\linewidth, trim=3.2cm 0cm 0cm 0cm]{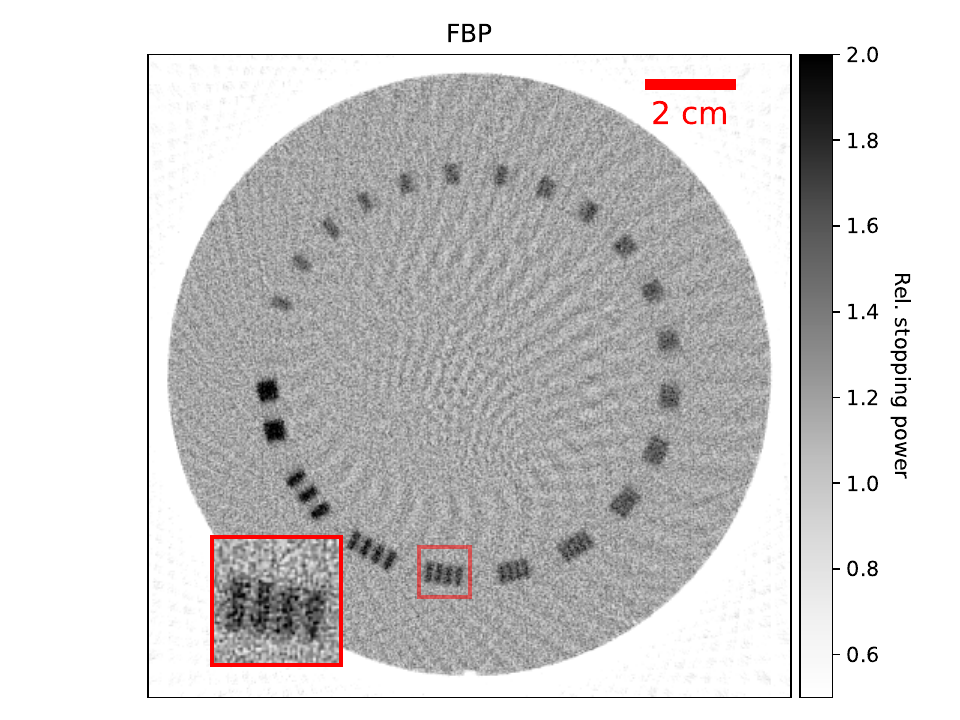}
    \includegraphics[width=0.32\linewidth, trim=3.2cm 0cm 0cm 0cm]{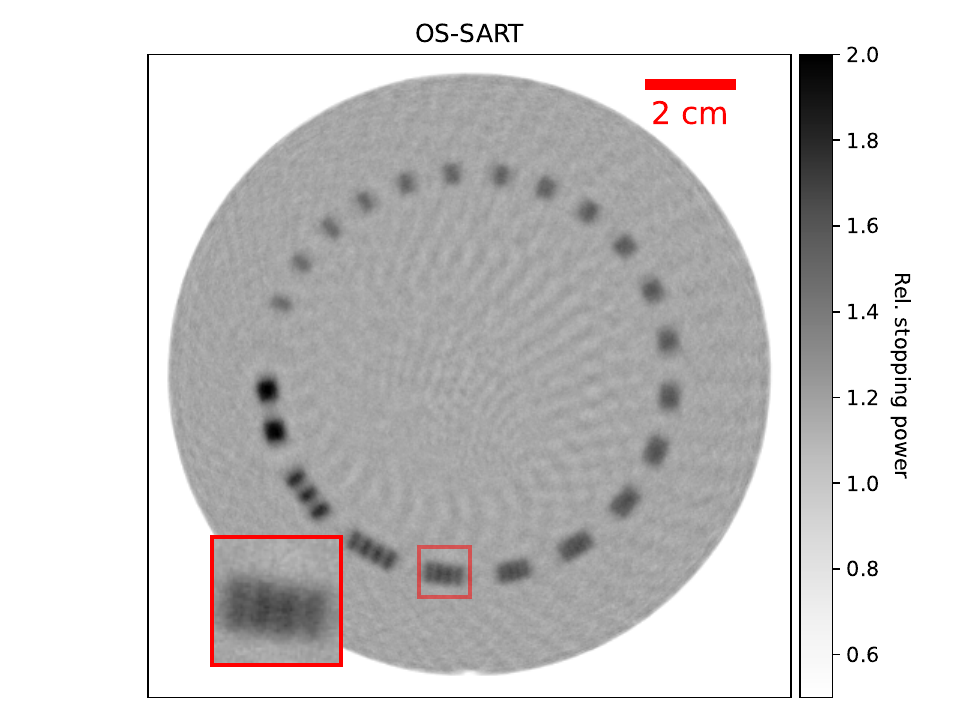}
    \includegraphics[width=0.32\linewidth, trim=3.2cm 0cm 0cm 0cm]{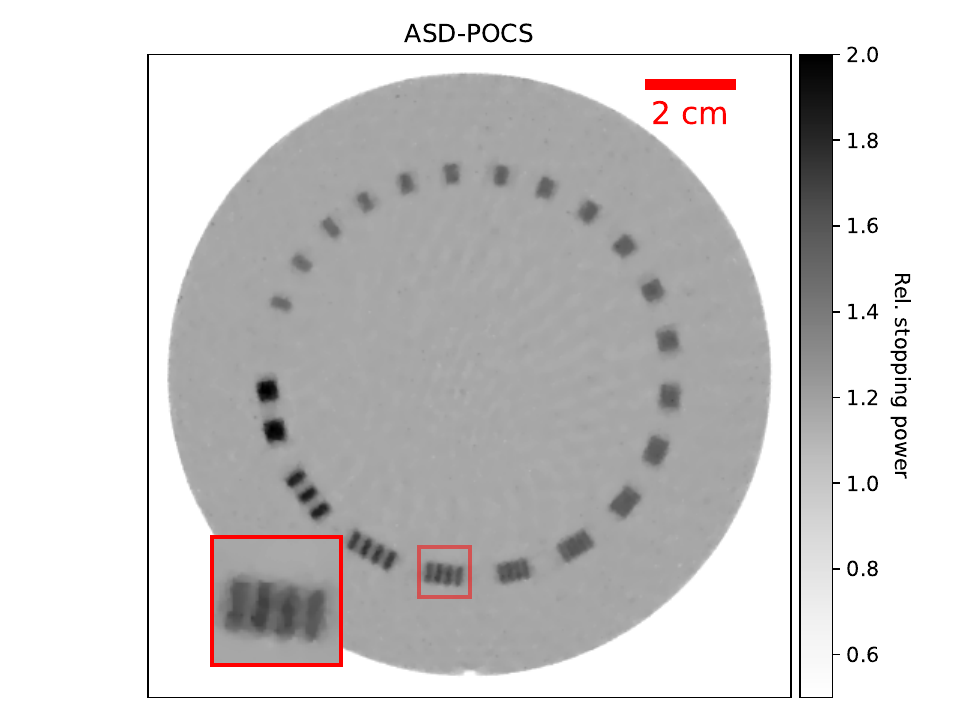}
    \caption{Reconstructions of pCT projections of the Catphan high resolution phantom \protect\cite{ctp_hr}.  Left: filtered-backprojection, middle: OS-SART, right: ASD-POCS.}
    \label{fig:pCT_recos}
\end{figure}

The resulting reconstructions are displayed in Figure~\ref{fig:pCT_recos}, for FBP, OS-SART at 100 iterations and ASD-POCS at 20 iterations. While the filtered-backprojection results in the noisiest image, OS-SART results in a smoother but blurrier image. In all cases, despite the very low number of projections used, the algorithms still allow viewing details of the aluminium line pair insets of the phantom. The least amount of noise in the reconstruction was obtained with the ASD-POCS algorithm.

\FloatBarrier
\subsection{Neutron Tomography dataset}

Neutron imaging is a highly accurate non-invasive method which can be used to image objects containing light elements encased inside a heavier material \cite{lohvithee2022NTreview}. This is because the properties of neutrons are complementary to those of X-rays, i.e. the transmissivity of neutrons with metallic elements is higher while their transmissivity with light elements is lower. However, this imaging technique sometimes leads to very slow measuring times, often making it impractical to obtain full angle measurements. For more information, we refer the reader to \ref{Appendix12}.

The neutron image dataset in this experiment was obtained at the Thai Research Reactor-modification 1 (TRR-1/M1) located at the Thailand Institute of Nuclear Technology (TINT). In this facility, it generally takes at least 14 hours to acquire the 501 projection images, covering 180 degrees in intervals of 0.36 degrees required to obtain a crisp reconstruction using FBP. This long scanning time is currently limiting their potential applications, so they are exploring the use of iterative algorithms to produce reconstructions of similar quality with a limited number of projections. 

\begin{figure}
    \centering
    \includegraphics[width=0.75\linewidth]{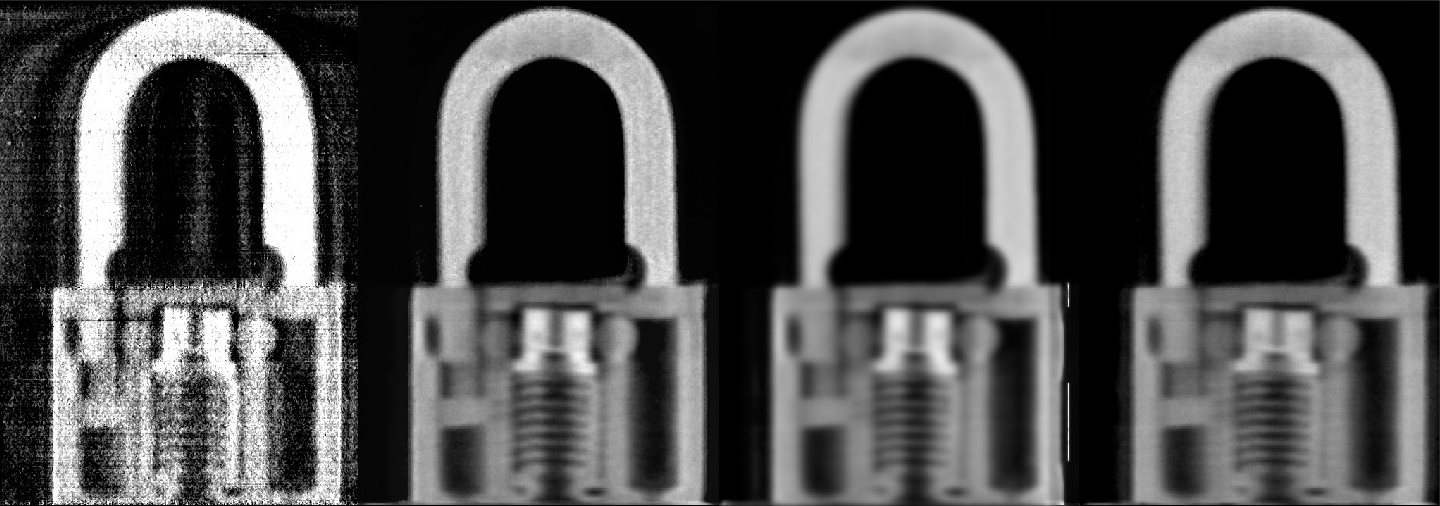}
    \caption{ Neutron tomography dataset of the padlock taken at Thailand Institute of Nuclear Technology (TINT). From left to right: reconstructions using FDK with Shepp-logan filter, SART, SART-TV and SIRT. All the iterative algorithms were executed at 100 iterations.}
    \label{fig:neutron}
\end{figure}

For the presented experiment, 
the thermal-neutron flux at the imaging position was 2.5 × 10$^5$ cm$^{-2}$s$^{-1}$. The main instruments of the neutron imaging system consist of a neutron-to-photon conversion plate made from 6LiF/ZnS, a 45-degree mirror, a sample rotation system operated by in-house developed software, a Nikkon 50-mm/f1.2 lens, and a 2048$\times$2048-pixel CCD camera. 
The Field-of-View (FOV) is $20$$\times$$20$ cm$^2$ and the L/D ratio is 78. 
In this experiment, four reconstruction algorithms (FDK, SART, SART-TV, and SIRT) were used to reconstruct the same neutron tomography dataset with an extremely limited set of neutron projections. The sizes of the projection data are 887$\times$887 pixels, taken on an arc range of 180$^{\circ}$ in 18$^{\circ}$ increments, leading to a total of 10 projection images per set (as opposed to the standard 0.9$^{\circ}$ increments, leading to a total of 201 projection images per set). The size of the reconstructions is 512$\times$512$\times$720 voxels. 
The filter Shepp-logan was used for the FDK reconstruction. One hundred iterations were executed for the SART, SART-TV, and SIRT algorithms. The codes that produce the reconstructed images in Figure \ref{fig:neutron} can be seen in Snippet \ref{code:neutron}.

\begin{minipage}{\linewidth} 
\begin{lstlisting}[language=Matlab,caption=MATLAB code to reconstruct neutron tomography data,label=code:neutron]
% Load TIGRE
InitTIGRE;
% Load data
[proj,geo,angles] = TINTDataLoader(data_folder) % This function is not in TIGRE
% Reconstruction
recon_fdk_shepp_logan = FBP(proj, geo, angles,'filter','shepp-logan');
recon_sart            = SART(proj,geo,angles,100); %SART for 100 iterations
recon_sirt            = SIRT(proj,geo,angles,100); %SIRT for 100 iterations
recon_sart_tv         = SART_TV(proj,geo,angles,100); %SART_TV for 100 iterations
\end{lstlisting}
\end{minipage}

A representation of the reconstructed images is shown in Figure \ref{fig:neutron}: the left-most column shows the SIRT reconstruction; the second column shows the SART\_TV algorithm, which is based on SART with additional TV regularization; the third column shows the standard SART reconstruction; and the right-most column shows the FDK reconstruction with a Shepp-logan filter. Since this experiment considered an extremely limited set of data, the resulting images lack definition around the edges of the padlock components in all reconstructions. However, it can be obviously seen that the results obtained using the FDK algorithm show more severe artifacts. If we compare the results obtained using the three iterative algorithms, we can observe in the top part of the padlock on column three of Figure \ref{fig:neutron} that the reconstruction given by SART preserves edges better than the other algorithms.

\subsection{Diondo \textmu CT dataset}
When radiation dose is not an issue, micrometre-range pixel resolution \textmu CT scanners are often used, particularly increasingly being used in industry for non destructive testing. The high resolution, however, is linked to long scanning times; and while this is acceptable for some applications, it is a hindrance if the scanning is used for continuous inspection of samples. In \textmu CT, when there is no limit to the scanning time and if we have access to the object from all directions and if the object is approximately cylindrical with no significant variations in X-ray path length through the object, the FDK algorithm produces very high quality reconstructions: with enough radiation exposure, the signal to noise ratio of the data is sufficiently high to minimally impact the reconstruction. This is not true when faster scans are needed, as either the projection data has to be acquired faster, thus with more noise, or less projection data are acquired. This example falls into the latter category where, to speed-up measuring times, an Energizer AAAA battery is measured in only a subset of 258 out of the 1258 projection angles prescribed in a conventional scanning regime. The scan was acquired at the University of Southampton \textmu Vis lab, which is part of the National X-ray Computed Tomography (NXCT) facilities of the UK, on a diondo d5 using a 300 kVp X-ray transmission source, with a detector size of $2000 \times 2000$ (reduced by selecting the $800 \times 2000$ pixel central region), in helical trajectory mode. 

\begin{figure}
    \centering
    \includegraphics[width=0.8\linewidth]{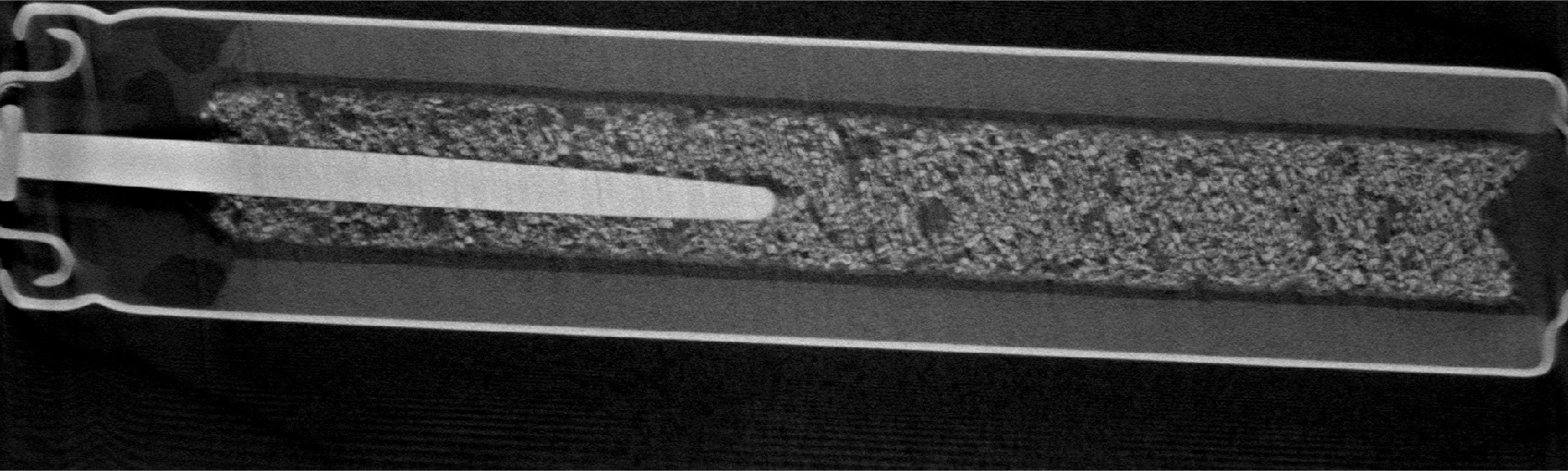}
    \includegraphics[width=0.8\linewidth]{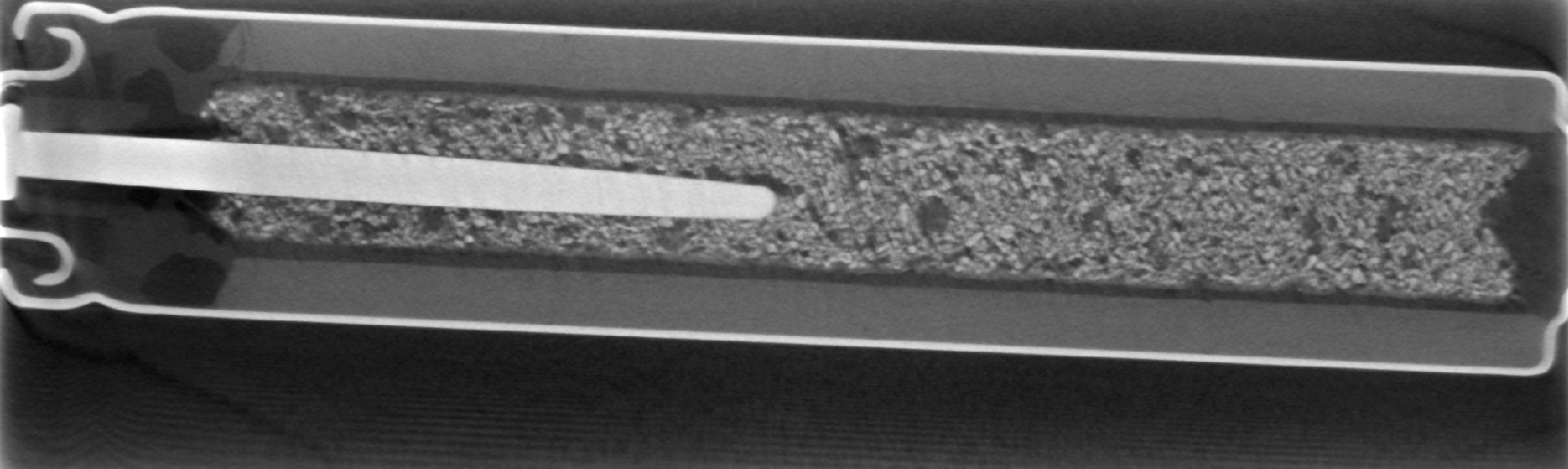}
    \caption{Scan of an Energizer AAAA battery acquired in a diondo d5 scanner at the \textmu Vis lab at the University of Southampton using a helical acquisition mode. Top shows FDK and bottom FISTA with TV regularization (20 iterations). }
    \label{fig:battery}
\end{figure}

A representation of the reconstructions obtained using FDK and the FISTA algorithm with TV regularization run over 20 iterations can be seen in Figure \ref{fig:battery}, following the code in Snippet \ref{code:battery}. The images are $600 \times 600 \times 2000$ voxels. The reconstruction obtained using FISTA seems to be much less noisy, albeit a bit blurry in some areas. A better choice of hyperparameters may be able to improve the reconstruction, however it is also important to note that FDK is sometimes sufficiently performing, and this may be one of such cases. It is likely that FISTA would be a better algorithm to use for segmentation, but if visual inspection is only needed, FDK does a perfectly fine job. This is ultimately true for any reconstruction algorithm and application: image quality is dependent on what the image is needed for, it is not itself a target.

\begin{minipage}{\linewidth} 
\begin{lstlisting}[language=Matlab,caption=MATLAB code to reconstruct Diondo Helical data,label=code:battery]
% Load TIGRE
InitTIGRE;
% Load
[projections, geo, angles] = DiondoDataLoader("./path/to/battery/scan");
% Reconstruct
recon_fdk   = FDK(projections,geo,angles);
recon_fista = FISTA(proj,geo,angles,20,'hyper',2e2,'init','FDK','tviter',50);
\end{lstlisting} 
\end{minipage}

\section{Discussion and Conclusions}\label{Sec:conclusions}

In this work, we present nearly a decade’s worth of scientific advancements in the field of CT reconstruction, encapsulated in the open-source code contributions to the TIGRE toolbox. These contributions correspond to a twenty-fold increase in the code-base of the toolbox, and have allowed an effective and practical transfer of knowledge from the mathematical fields of optimization and inverse problems into the experimentalists' world, as the various examples in this article, and hundreds of users of TIGRE show. 

One of the main goals of the TIGRE toolbox is to provide tested and robust implementations of iterative algorithms that can be used on real data acquisitions in lab-environments. Moreover, whenever possible, the released implementations have been standardized (in terms of parameters choices, etc) to ease comparison and applicability. Currently, it already contains 23 standardized algorithms, and this number keeps increasing.

As highlighted in this article, stand-alone reconstruction algorithms are not sufficient for reconstructing real data. On the one hand, effective CT reconstruction requires operators (forward and backward) that are not only geometrically adaptable but also computationally efficient, and that can be incorporated easily in any structured algorithm implementation. Since its first release, TIGRE has kept on improving the computational speed and flexibility of its operators, facilitating the seamless reconstruction of experimental data. On the other hand, handling data from various scanners is a complex task involving thousands of lines of code, which is crucial both to simplify the testing of different algorithms by practitioners, as it is to successfully evaluate the performance of a new algorithm for different problems. To address this, TIGRE now includes data-loading support for seven different manufacturers, all of them verified on real scanners.

While great efforts to simplify the use of iterative algorithms for non-experts has been done, the authors acknowledge that it is sometimes complex to know which algorithm to use under specific scanning conditions or how to fine-tune them to the appropriate parameters.  Suggestions and guidelines exist in the TIGRE repository, but future work will rely on improving automatic parameter selection and attempting to have a more systematic way to help users select algorithms.

On the topic of algorithm selection, it is important to mention that defining what makes an algorithm good is not possible. It is not uncommon to promote algorithm performance assuming that the final image quality compared to a reference is the most important feature (even if the most commonly used metrics fail at this task \cite{breger2024study,breger2024study2}), however image quality itself is a relative metric. For example, an image for diagnosis in medical CT or an image for radiation treatment would focus on different things, detectability in the first case, but quantitative accuracy in the second. The best image reconstruction for each of these tasks is not the same. Similarly, a \textmu CT image for metrology or material characterization would be evaluated very differently on what constitutes quality. This is why it is so relevant that many iterative algorithms exist, and why it is important that these can be readily used.

In this work, we also want to highlight the broad range of applicability of iterative algorithms, and of the toolbox itself, with respect to different imaging types. We highlighted examples of all the main 3D CT modalities, namely synchrotron CT, medical CBCT and \textmu CT, and also neutron CT and proton-CT. The (forward) mathematical model behind the first four is based on the same operator, the Radon transform, but the way the data is handled and loaded is very different, as well as the nuisances of each application. Moreover, proton-CT, has also been adapted to fit seamlessly into the TIGRE framework. In this way, the experiments in this paper aim to showcase how TIGRE can be easily used in very different kinds of tomographic problems.

TIGRE has grown significantly since its inception, mostly driven by user feedback and needs, and it is the intention of the authors for this momentum to continue. Its original publication shows a snapshot of what the software was when it was released, and here we offer another snapshot almost a decade from then. Now, the purpose of this work is to provide a structured overview of the current version of the toolbox, providing appropriate descriptions and references; and to serve as a comprehensive and peer-reviewed guide for the users. It is, however, not the intention of this paper to provide a complete version of the toolbox, which we hope may become incomplete very soon with the addition of new capabilities.  
We thus urge mathematicians, CT experimentalist and anyone in-between to join into the TIGRE software community and contribute, from code, to issues, to feature requests at \url{github.com/CERN/TIGRE/}.

\section*{Acknowledgments}
A. Biguri wants to thank all named and anonymous contributors of TIGRE toolbox over the past years, for keeping TIGRE alive and updated with the latest technological trends. The software is alive thanks to all these contributions. A. Biguri would also like to acknowledge CT users that have come with interesting problems to solve and many bugs to fix, as without the requests of these users, TIGRE would not be able to accommodate so many real use-cases. A. Biguri will finally also like to acknowledge CERN, EPSRC (various grants) and the Accelerate Programme for
Scientific Discovery for funding that allowed to develop and maintain such tool.  The authors acknowledge the \textmu-VIS X-ray Imaging Centre (\url{muvis.org}), part of the National Facility for laboratory-based X-ray CT (\url{nxct.ac.uk} – EPSRC: EP/T02593X/1), at the University of Southampton for provision of tomographic imaging facilities.

This work was partially supported by the National Science Foundation program
under grant DMS-2208294. Any opinions, findings, and conclusions or recommendations expressed in this material are those of the
author(s) and do not necessarily reflect the views of the National Science Foundation.
\section*{Author Contribution}

A. Biguri is the main maintainer and developer of TIGRE and has been involved in every part of the work presented here. T. Sadakane and Y. Liu are active contributors to the tool and have significantly added to the general TIGRE development and issue fixing, particularly in the Python side. R. Lindroos developed the first iteration of the Python-TIGRE library. M. Sabaté Landman's contributions are related to the implementation of Krylov subspace methods, algorithm features and paper structure. M. Lohvithee developed total variation regularization algorithms and the experiments with neutron CT. S. Kaser created the proton CT methods. S. Hatamikia developed arbitrary rotation code and invaluable algorithmic testing expertise. Y Du developed, tested and validated the TIGRE module (TIGRE-Varian) for Varian linac onboard CBCT in clinical settings \cite{VARIAN}. R. Bryll significantly improved the speed of the GPU-driven operators. E. Valat implemented the PyTorch bindings. S. Wonglee set the neutron tomography experiments and acquisition. C.B. Sch{\"o}nlieb and T. Blumensath provided knowledge and support on reconstruction methods. 

\section*{Bibliography} 

\bibliographystyle{unsrt}
\bibliography{bib}

\newpage
\appendix
\section{Table of algorithms} \label{Appendix1}

\begin{table}[h!]
    \centering
        \footnotesize	

\begin{tabular}{ |l|c|c| } 
\hline
\bf Algorithm name & \bf Original  & \bf TIGRE  \\
& \bf description & \bf  description \\
  \hline
 \textbf{Direct methods} &  &  \\ 
 FBP and FDK & \cite{wang2002x}, \cite{wesarg2002parker}& \cite{TIGRE}\\
 \hline 
 \textbf{Kaczmarz-type algorithms} & & \\
 Simultaneous ART (SART) & \cite{SART} & \cite{TIGRE}\\
 Ordered Subsets ART (OS-SART) & \cite{OS_SART} & Section \ref{sec:OS}, \cite{TIGRE} \\
 Simultaneous Iterative Reconstruction Algorithm (SIRT) & \cite{SIRT} & \cite{TIGRE} \\
 \hline  
 \textbf{Projection Onto Convex Sets (POCS)} &&\\
 Adaptative Steepest Descend version (ASD-POCS) & \cite{ASD_POCS} &  \cite{TIGRE}, \cite{lohvithee2017parameter}\\
 Projection-Controlled Steepest Descent (PCSD) & \cite{liu2015reconstruction} & \cite{lohvithee2017parameter}\\
 Adaptive weighed TV norm & \cite{liu2012adaptive,lohvithee2017parameter} & Section \ref{sec:features}, \cite{lohvithee2017parameter} \\
 OS version with both TV norms &&  Section \ref{sec:OS}\\ 
 Bregman outer iterations + ASD-POCS (B-ASD-POCS-$\beta$) & \cite{B_ASD_POCS_beta} & \cite{TIGRE}\\
\hline  
 \textbf{Krylov Subspace algorithms} &&\\
 Conjugate Gradient Least Squares (CGLS) & \cite{hestenes1952methods} & \cite{TIGRE}, \cite{KrylovTigre2023}\\
 LSQR & \cite{paige1982algorithm} & \cite{KrylovTigre2023}\\
 hybrid LSQR & \cite{paige1982algorithm} & \cite{KrylovTigre2023}\\
 AB/BA-GMRES & \cite{hansen2022gmres} & \cite{KrylovTigre2023}\\
 LSMR & \cite{fong2011lsmr} & \cite{KrylovTigre2023}\\
 IRN-TV-CGLS & \cite{wohlberg2007iteratively} & \cite{KrylovTigre2023}\\
 hybrid-fLSQR-TV &\cite{gazzola2019flexible} & \cite{KrylovTigre2023}\\
  \hline 
  \textbf{Statistical minimization algorithms} & &\\
  OSEM &\cite{hudson1994accelerated} & Section \ref{sec:tomo_iter}\\
  \hline 
  \textbf{Proximal algorithms} & &\\
  Fast Iterative Shrinkage-Thresholding Algorithm (FISTA) & \cite{beck2009fast}& Section \ref{sec:tomo_iter}\\
  Fast variant of FISTA & \cite{liang2018faster} & Section \ref{sec:tomo_iter}\\
  SART-TV & \cite{SART_TV}  & \cite{TIGRE}\\
  \hline
\end{tabular}
    \caption{Here we provide a list of all the algorithms provided in the TIGRE toolbox (as of November 2024), along with the appropriate citations to the original papers, and the references that are specific to the TIGRE toolbox. Note that \cite{TIGRE} refers to the first publication of the toolbox.}
    \label{tab:my_label}
\end{table}

\section{Geometric Parameters in TIGRE}\label{Appendix_geo}
\begin{table}[h]
    \footnotesize	
    \centering
    \begin{tabular}{|p{20mm}| p{80mm} | p{15mm}|}
        \hline

        Geometric Parameter & Description & Reference\\
        \hline \hline
        DSD & Distance between the source (S) and the detector (D) & - \\ \hline
        DSO & Distance between the source (S) and origin (O) & -\\ \hline
        sVoxel & Size of the volume ($\mathbb{I}$) in real world units & - \\ \hline
        nVoxel & Number of voxels of the discretization of the $\mathbb{I}$ & - \\ \hline
        sVoxel & Size of each voxel in the discretization of $\mathbb{I}$& - \\ \hline
        offOrigin ($\vec{V_{orig}}$) & Shift of location of desired center of the volume w.r.t. the default & Oxyz\\ \hline
        COR & Center of Rotation shift. Moves the rotation axis & Oy\\ \hline
        angles $(z_1,y_1,z_2)$ & Euler ZYZ rotation angles & Oxyz\\ \hline
        sDetector & Size of the detector ($\mathbb{D}$) in real world units & - \\ \hline
        nDetector & Number of pixel of the discretization of the $\mathbb{D}$ & - \\ \hline
        sDetector & Size of each pixel in the discretization of $\mathbb{D}$& - \\ \hline
        offDetector ($\vec{V_{det}}$) & Shift of location of desired center of the detector w.r.t. the default & uv\\ \hline
        rotDetector $(\phi,\theta,\psi)$ & In place rotations of the detector & D\\
        \hline

    \end{tabular}
    \caption{Geometry parameters of TIGRE and description.}
    \label{tab:geometry}
    
\end{table}

\FloatBarrier
\section{Neutron Tomography}\label{Appendix12}


Neutron imaging is a highly accurate and reliable non-invasive measurement method that has been utilized for various applications such as cultural heritage, thermal-hydraulics, nuclear engineering studies, and botanical sciences \cite{lohvithee2022NTreview}. The properties of neutrons are complementary to those of X-rays and gamma-rays i.e. the transmissivity of neutrons with metallic elements is higher while their transmissivity with light elements is lower. Therefore, neutron imaging is excellent for inspecting specimens containing light elements inside the object to be scanned, especially if they are covered or enveloped by heavy elements on the outside.


Despite the advantages of neutron tomography, this modality has not yet been widely employed for routine inspection in industrial applications. One of the main reasons for this is the requirement of having a high neutron flux to produce high quality neutron tomography imaging, which has to be generated in a nuclear reactor or in a spallation source and is therefore not always easily accessible. Moreover, some common challenges can affect the power at which the reactors operate, as well as some potential benefits in terms of safety and money saving, causing a low neutron flux to arrive at the neutron imaging facility and resulting in very slow measuring times. This is important as neutrons also lead to the objects being 'activated' i.e. they become radioactive.
 In practice, the scanning time required for a full angle measurement can easily become impractical and compromise the imaging of many dynamic processes. 
Iterative algorithms have been shown to produce reconstructions with superior quality using a limited number of projections with respect to the commonly used FBP algorithm. 
Therefore, it is a promising avenue to explore the use of these methods in this context, 
with the aim of reducing the number of required projections and shorten acquisition times at neutron imaging facilities. 

An example of neutron imaging facility is the Thai Research Reactor-modification 1 (TRR-1/M1), located at the Thailand Institute of Nuclear Technology (TINT). The TRR-1/M1 research reactor is an open pool type of TRIGA-Mark III, whose main purpose is public research and non-destructive investigation using radiography and tomography techniques. 
This reactor has a limited resource of fuel rods, which employ old, discontinued technology. Hence, the reactor has to operate at 1 MW to save fuel. In this facility, the acquisition time required to obtain a good-quality projection for one particular angle takes approximately 100 seconds. This long scanning time is currently limiting their potential applications.

\end{document}